\documentclass[sigconf]{acmart}
\usepackage{subfig}
\usepackage{multirow}
\AtBeginDocument{%
  }

\setcopyright{acmlicensed}
\copyrightyear{2018}
\acmYear{2018}
\acmDOI{XXXXXXX.XXXXXXX}
\acmConference[Conference acronym 'XX]{Make sure to enter the correct
  conference title from your rights confirmation email}{June 03--05,
  2018}{Woodstock, NY}
\acmISBN{978-1-4503-XXXX-X/2018/06}

\begin{document}

\title[CANote: Empowering Fact-checking Note Writing\\ Through Scaffolded and Provenance-based Human-AI Collaboration]{CANote: Empowering Fact-checking Note Writing Through Scaffolded and Provenance-based Human-AI Collaboration}

\author{Shuning Zhang}
\authornotemark[1]
\email{zsn23@mails.tsinghua.edu.cn}
\affiliation{%
  \institution{Tsinghua University}
  \city{Beijing}
  \country{China}
}

\author{Jingruo Chen}
\authornote{Equal contribution.}
\email{jc3564@cornell.edu}
\affiliation{%
  \institution{Information Science, Cornell University}
  \city{Ithaca}
  \state{New York}
  \country{U.S.}
}

\author{Yuwei Chuai}
\authornotemark[2]
\email{yuwei.chuai@uni.lu}
\affiliation{
    \institution{SnT, University of Luxembourg}
    \city{Esch-sur-Alzette}
    \country{Luxembourg}
}

\author{Dai Shi}
\email{shidai@tongji.edu.cn}
\affiliation{
    \institution{College of Design and Innovation, Tongji University}
    \city{Shanghai}
    \country{China}
}

\author{Yifan Wang}
\email{yifanw57@uw.edu}
\affiliation{
    \institution{University of Washington}
    \city{Seattle}
    \state{Washington}
    \country{U.S.}
}

\author{Xin Yi}
\authornote{Corresponding authors.}
\email{yixin@tsinghua.edu.cn}
\affiliation{
    \institution{Tsinghua University}
    \city{Beijing}
    \country{China}
}

\author{Hewu Li}
\email{lihewu@cernet.edu.cn}
\affiliation{
    \institution{Tsinghua University}
    \city{Beijing}
    \country{China}
}

\renewcommand{\shortauthors}{Trovato et al.}

\begin{abstract}
  Crowdsourced fact-checking mechanisms, such as $\mathbb{X}$'s Community Notes, play a critical role in mitigating the spread of misinformation. However, drafting high-quality, evidence-based debunking notes imposes a substantial burden on contributors. We present CANote, an AI-assisted debunking note writing system featuring evidence correlation and structured co-drafting. CANote scaffolds the workflow by extracting subclaims from social media posts, providing provenance through explicit links between subclaims and retrieved evidence, and generating neutral, structural drafts to support human reasoning. We evaluated CANote against manual writing (N=52 fact-checkers, N=52 lay users) on simulated $\mathbb{X}$ platform, where we found CANote significantly improves note quality. Notably, CANote enables lay users to write notes that have comparable quality to those written by experts. While the task completion time and perceived cognitive load remain comparable to manual drafting, CANote significantly increases user satisfaction. However, this assistance introduces a trade-off, resulting in a reduced sense of user ownership and control over the debunking note. \textcolor{red}{Warning: this work contains misleading content that some may find distressing.}
\end{abstract}


\keywords{Do, Not, Use, This, Code, Put, the, Correct, Terms, for,
  Your, Paper}
\begin{teaserfigure}
    \centering
  \includegraphics[width=\textwidth]{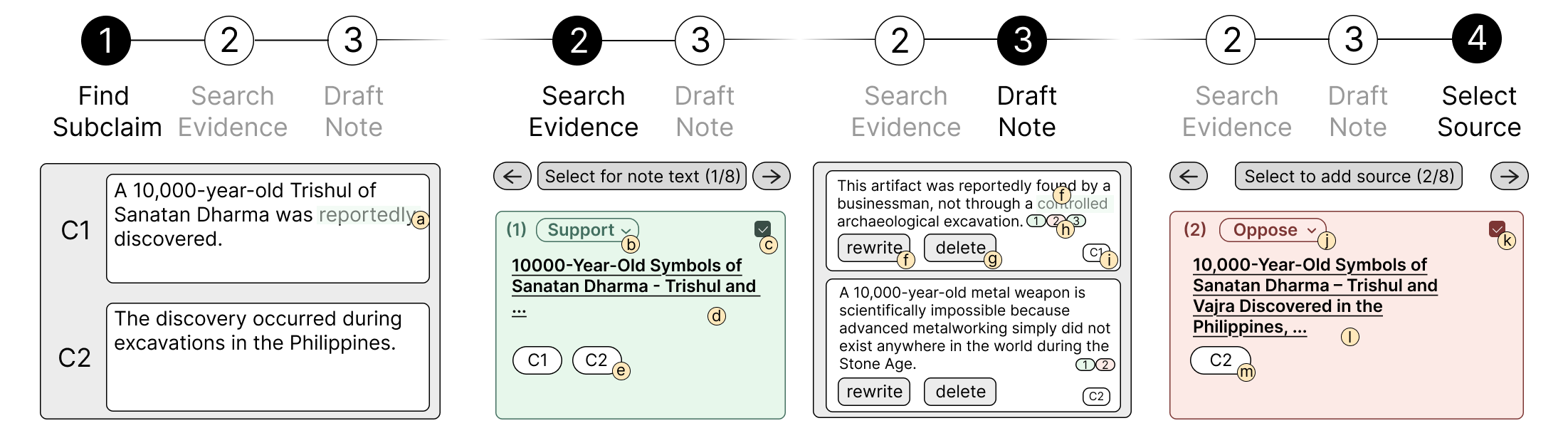}
  \caption{The CANote workflow, which features a four-stage human-in-the-loop pipeline for AI-assisted debunking note writing. (1) Find Subclaim: decomposes complex claims into verifiable units (e.g., C1, C2) where \textit{users may \textcircled{a} edit each subclaim}. (2) Evidence Retrieval: provides relevant information and enables \textit{users to \textcircled{b} adjust stances, \textcircled{c} toggle inclusion, \textcircled{d} inspect evidence, and \textcircled{e} trace subclaim links}. (3) Collaborative drafting: generates stance-based summaries that users can \textit{\textcircled{f} edit or \textcircled{g} remove while verifying links to \textcircled{h} sources and \textcircled{i} subclaims}. (4) Source selection: finalizes the note evidence through \textit{\textcircled{j} stance correction and \textcircled{k} inclusion management, ensuring transparency via \textcircled{l} evidence and \textcircled{m} subclaim cross-references}.}
  \Description{Illustration of CANote, an explainable, proactive AI-assisted note writing system with a four-stage human-in-the-loop process. Evidence items are labeled as supporting or opposing claims, linked to subclaims through clickable indices, and synthesized to produce structured note text. This staged design operationalizes verification as part of the normal interaction flow, enabling users to iteratively inspect, edit, and validate claim-level support before submission, thereby improving transparency, user agency, and auditability of AI-assisted note writing.}
  \label{fig:teaser}
\end{teaserfigure}


\maketitle

\section{Introduction}
The rapid spread of misleading content, misinformation/disinformation online necessitates scalable and rapid verification mechanisms. Crowdsourced fact-checking platforms, most notably $\mathbb{X}'s$ Community Notes, have emerged as a primary solution by leveraging collective intelligence to append critical context directly to misleading posts~\cite{chuai2024did,chuai2024community}. While effective, the manual process of contributing a note is highly demanding~\cite{warren2025show}. Contributors must identify checkable claims, search across external sources to verify facts, synthesize multiple streams of information, and draft a concise response that adheres to neutrality guidelines. This multi-step process introduces high cognitive load, deterring potential contributions and causes critical delays, allowing misinformation to spread during the critical window before context is appended~\cite{chuai2024community,zhang2025commenotes}.

While prior mixed initiative systems~\cite{nguyen2018believe,fu2024data} and platforms like $\mathbb{X}$ have explored AI assisted note writing, these efforts remain constrained~\cite{communitynotes_collaborative_notes}. Current implementations typically restrict collaboration to AI generating drafts and humans providing text based feedback. This integration lacks explainability and interleaved interaction, severely limiting the capability of human value judgment and AI evidence retrieval. Furthermore, recent studies detailing user perspectives on AI fact checking tools have yet to be translated to concrete interactive systems~\cite{liu2024human,warren2025show}.

To address these challenges, we propose CANote, an AI-assisted debunking note contribution technique. CANote scaffolds the human fact-checker through a provenance-based explainable pipeline. Recognizing that information gathering and unbiased synthesis are the primary challenges~\cite{warren2025show}, our technical design focuses on two core aspects:

(1) Provenance-based evidence retrieval: To move beyond black-box fact-checking, CANote provides direct provenance between subclaims and retrieved evidence. By leveraging similarity ranking, the system generates visual linkages that allow users to trace the specific verification evidence for each subclaim. These mappings are augmented by a mixed-stance visualization that intentionally surfaces both supporting and opposing evidence, paired with concise, phrase-level rationales. By externalizing these underlying correlations through multi-perspective links, CANote reallocates the contributor's cognitive effort from information gathering toward critical auditing, thereby mitigating automation bias.

(2) Scaffolded co-drafting: CANote decomposes the debunking note writing process into a four-stage pipeline~\cite{wang2024factcheck}, including find subclaim, search evidence, draft note, and select source. CANote connects each stage with evidence, subclaims and note text, ensuring that human interventions at any stage propagate throughout the workflow. Furthermore, the interface embeds clickable, grouped citation markers into the editable note draft, enabling users to trace specific sentences back to their source subclaims and evidence IDs. This high-agency design allows contributors to treat AI draft as a verifiable scaffold rather than an opaque final output.

We validated our design through a between-subjects user study (N=52 fact-checkers, N=52 lay users), evaluating CANote against a manual baseline mimicking the current Community Notes workflow. We found CANote significantly improves overall note quality without increasing drafting time. Furthermore, CANote effectively enables lay users to write notes that match expert standards. While overall cognitive load remained comparable to manual drafting, participants rated CANote as significantly more supportive and efficient. Notably, we also found a diminished control on the notes when participants used CANote. This warrants future exploration.

Collectively, we make the following contributions:

$\bullet$ We present CANote, an AI-assisted debunking note writing system that integrates provenance-based evidence retrieval with transparent co-drafting to support fact-checkers.

$\bullet$ Our evaluation shows that CANote significantly improves note quality and enables lay users to match expert performance, without increasing task time or cognitive load.

$\bullet$ We identify trade-offs in human-AI collaboration, revealing that while the technique increases user satisfaction, it diminishes perceived ownership.




\section{Related Work}

\subsection{AI-Assisted Fact-Checking}

Researchers have increasingly explored AI to assist human fact-checking workflows. We review prior literature across two primary areas: developing systems to identify and classify claims, and designing explainable interfaces to support human reasoning.

\textbf{Claim detection and classification.} To handle the growing volume of misinformation, prior systems have focused on automating the detection and triage of claims. Approaches include unsupervised pipelines for clustering redundant claims~\cite{yang2021scalable}, hybrid frameworks combining AI and crowdsourcing~\cite{la2022hybrid}, and customizable LLM filters to prioritize checkworthy content~\cite{liu2025exploring}. Researchers also emphasize co-designing these Natural Language Processing (NLP) tools to align with professional fact-checkers' practices~\cite{liu2024human}. Other interactive tools assist verification by calculating veracity scores based on adjustable source reputations~\cite{nguyen2018believe} or by mapping text to verify claims against structured datasets~\cite{fu2024data}.

However, these systems primarily focus on the early stages of the fact-checking pipeline, specifically automated classification and structured data verification. In contrast, CANote also supports the downstream generative workflow. Rather than solely estimating the probability of a claim being true~\cite{nguyen2018believe} or relying on structured datasets~\cite{fu2024data}, CANote scaffolds the human-in-the-loop pipeline to navigate unstructured, conflicting web evidence and co-drafting debunking notes.

\textbf{Explainability and user interaction.} Beyond detection, systems are designed to  explain their reasoning, supporting human judgment~\cite{zhang2026collab}. Basic Explainable AI (XAI) and simple warning labels can reduce intentions to share misinformation~\cite{lim2023xai,epstein2022explanations}. Furthermore, personalized AI predictions~\cite{jahanbakhsh2023exploring} and disclosing an AI's identity~\cite{lan2025ai} can influence users' reliance and trust. Yet, the use of generative AI in knowledge work introduces a risk of over-reliance by shifting the user's focus heavily toward verification~\cite{lee2025impact}. Professional fact-checkers, in particular, require tools that go beyond basic labels. They need systems that trace logical reasoning, cite explicit evidence, and articulate uncertainty~\cite{warren2025show}.

Current XAI systems address this by generating post-hoc explanations for automated classifications~\cite{zhang2021faxplainac,vargas2024improving} or providing dashboards to evaluate how XAI features affect user trust~\cite{schmitt2024evaluating}. We advance these research by positioning explainability as a generative scaffold rather than a diagnostic tool. Unlike systems designed to evaluate trust~\cite{schmitt2024evaluating,lim2023xai} or explain binary classifications~\cite{zhang2021faxplainac,vargas2024improving}, CANote provides provenance-based evidence correlation to scaffold the actual creation process, decomposing complex claims and co-drafting evidence-based responses.

\subsection{Expert Fact-Check Flow}

The fact-checking landscape has evolved from internal journalistic verification into a complex ecosystem of professional, automated, and crowdsourced efforts. Foundational studies established the origins and methodologies of this practice~\cite{graves2019fact}, proposing frameworks that navigate epistemological challenges, such as objectivism, through rigorous, transparent, and engaging presentation~\cite{suomalainen2025fact}. Complementing these studies, prior research has extensively investigated the work practices of fact-checkers~\cite{juneja2022human,micallef2022true} , revealing methodological nuances in statement selection and rating~\cite{markowitz2023cross}. Despite these advances, professional fact-checkers report unmet technological needs~\cite{dierickx2023journalism} and a misalignment between AI capabilities and expert workflows, highlighting a demand for human-centered tools with traceable reasoning~\cite{warren2025show}. Informed by these unmet needs, CANote explicitly embeds evidence within the drafting process.

To meet these workflow demands, automated fact-checking frameworks have evolved from basic NLP-driven claim detection~\cite{hassan2017toward} to sophisticated multi-stage LLM pipelines that emulate expert claim decomposition and synthesis~\cite{nair2025multi}. Researchers continually assess model reliability, mitigating hallucinations via retrieval-augmented generation (RAG)~\cite{rahman2026hallucination} and utilizing weighted scorers for retrieval accuracy~\cite{akhtar2024ev2r}. This evaluation ecosystem increasingly integrates large-scale knowledge graphs for complex reasoning~\cite{kim2023factkg}. Furthermore, modern systems support multimodal and multilingual verification, including zero-shot text-image pipelines~\cite{braun2024defame}, visual question-answering benchmarks~\cite{jiang2026pixels}, and low-resource language datasets~\cite{halitaj2025towards}. While these autonomous pipelines have inspired CANote's modular architecture, CANote prioritizes transparent design for human-in-the-loop debunking note drafting.

Fact-checking methodologies have also been tailored for high-stakes, specialized domains. In biomedical contexts, frameworks combine scientific evidence retrieval with LLM reasoning to validate claims~\cite{barone2025combining,bayani2025transformer}, compensating for human tendencies to judge health information based on superficial criteria~\cite{patterson2025professionally}. In the legal domain, specialized AI tools reduce errors but remain prone to hallucinations~\cite{magesh2025hallucination}. This vulnerability necessitates frameworks that address ethical challenges~\cite{koenecke2025tasks} and incorporate human-centric visual analytics to navigate legal complexity~\cite{furst2025challenges}. Similar hybrid models merging text search with statistical techniques have advanced the verification of historical statements~\cite{kobayashi2017automated}. While these domain-specific adaptations underscore the risks of unchecked AI in critical scenarios, CANote addresses such concerns through a transparent approach that empowers users to verify evidence and maintain control over the final note. 

Finally, crowdsourced fact-checking leverages collective intelligence to scale moderation efforts~\cite{chuai2024community,chuai2024did}. Analyses of platforms like $\mathbb{X}$'s Community Notes reveal that crowds can match expert efficiency in claim selection, yet they remain susceptible to partisan bias during source evaluation~\cite{saeed2022crowdsourced}. Comparative research indicates that machine learning and crowds with high political knowledge improve accuracy but fail to consistently meet professional standards~\cite{godel2021moderating}. This gap is frequently attributed to individual cognitive biases, including the affect heuristic and overconfidence, which distort the truthfulness judgments of non-expert workers~\cite{draws2022effects}. Motivated by these challenges, CANote introduces structured scaffolding to mitigate such discrepancies.

\section{CANote}\label{sec:CANote}

\subsection{Design Goals}

Based on prior guidance~\cite{warren2025show,liu2024human,juneja2022human}, we formulate three design goals:

\textbf{DG1. Scaffold the fact-checking process.} To reduce cognitive burden, the system should provide structural guidance throughout the complex, multi-step fact-checking workflow, supporting user creation without prescribing final outcomes~\cite{wang2024factcheck,nakov2021automated}.

\textbf{DG2. Establish transparent linking between content and claims.} To address the opacity of prior systems~\cite{schmitt2024evaluating,warren2025show,nguyen2018believe}, which can impede collaboration~\cite{warren2025show}, the system should present direct provenance through explicit mappings between claims and evidence to foster understanding and informed collaboration~\cite{warren2025show}.

\textbf{DG3. Present balanced evidence and enable free-form editing.} To mitigate conformity bias, political bias, and other LLM biases~\cite{moscovici1972social,bakke2025fact}, the system should present multiple sides of evidence and allow fact-checkers to freely edit all content, ensuring human judgment remains critical.

\subsection{Interaction Flow}

We show CANote's four-stage interaction flow through a typical usage scenario. Consider Bob, a volunteer fact-checker reviewing a complex viral post claiming that a 10,000-year-old Trishul of Sanatan Dharma was discovered during excavations in the Philippines.

\begin{figure}[h]
    \centering 
    \includegraphics[width=0.47\textwidth]{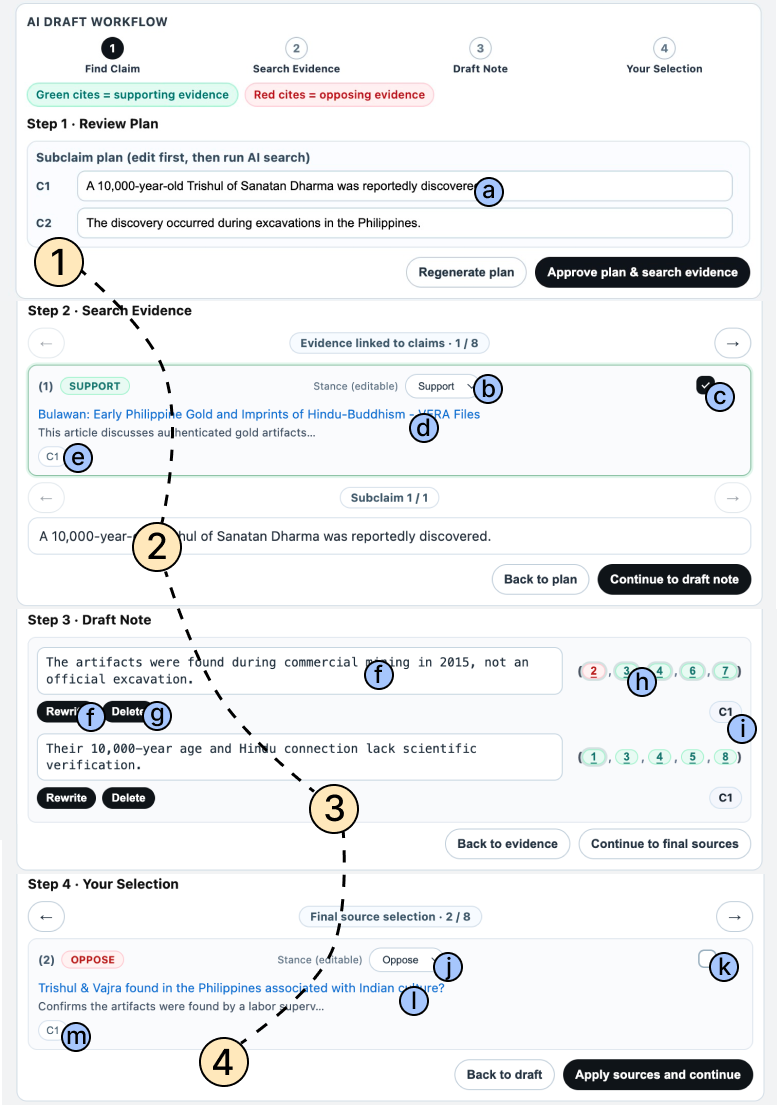}
    \caption{The interface of CANote.}
    \label{fig:canote_interface}
\end{figure}

In the first stage (Fig~\ref{fig:canote_interface} \textcircled{1}), CANote finds two subclaims C1 and C2 in the post. Bob reviews these generated subclaims, and may edit C1 to ensure the subclaim is appropriate before proceeding.

Next, CANote retrieves relevant information for subclaims, presenting horizontally scrollable evidence cards with AI-generated stance detection (i.e., green for ``Support'' and red for ``Oppose'') (Fig~\ref{fig:canote_interface} \textcircled{2}). Bob clicks on the evidence (Fig~\ref{fig:canote_interface} \textcircled{d}) and reviews its alignment with the corresponding subclaim (Fig~\ref{fig:canote_interface} \textcircled{e}). Bob could change the stance of evidence (Fig~\ref{fig:canote_interface} \textcircled{b}) to ``Oppose'' or toggle the checkbox to determine whether to include this evidence in note drafting (Fig~\ref{fig:canote_interface} \textcircled{c}). This affects the note drafting results. 

\begin{figure*}[!htbp]
    \includegraphics[width=\textwidth]{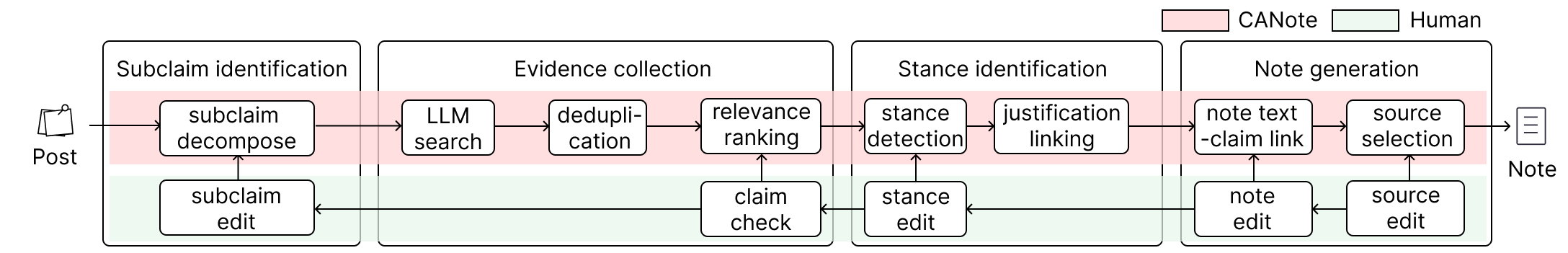}
    \caption{System flow of CANote, illustrating the iterative human-in-the-loop pipeline.}
    \label{fig:system-flow}
\end{figure*}

Once the evidence is curated, Bob enters the draft note stage (Fig~\ref{fig:canote_interface} \textcircled{3}). Here, CANote synthesizes the selected evidence into an editable draft. CANote displays clickable, grouped inline citation markers for each sentence (e.g., \textcircled{2}\textcircled{3}\textcircled{4} for the first sentence). Bob can click rewrite to edit the note (Fig~\ref{fig:canote_interface} \textcircled{f}), or delete (Fig~\ref{fig:canote_interface} \textcircled{g}) parts of the note. Bob can also click on citation index (Fig~\ref{fig:canote_interface} \textcircled{h}) or subclaim tag (Fig~\ref{fig:canote_interface} \textcircled{i}) to trace the generated text back to its source. 

Finally, Bob chooses the supporting or opposing evidence that will be directly attached to the published note as a source evidence (Fig~\ref{fig:canote_interface} \textcircled{4}). Similar to the evidence search phase, Bob verifies the evidence text (Fig~\ref{fig:canote_interface} \textcircled{l}), the linked subclaim (Fig~\ref{fig:canote_interface} \textcircled{m}), and corrects the stance if necessary (e.g., to Support for the second evidence, Fig~\ref{fig:canote_interface} \textcircled{j}).

\subsection{System Design}

To address the limited adoption of black-box AI fact-checking systems that fail to align with the complexities of professional workflows, CANote employs a human-in-the-loop framework centered on procedural scaffolding (DG1) and explicit provenance (DG2). Rather than providing opaque, automated verdicts, CANote scaffolds human-AI collaboration across a structured, four-stage workflow: subclaim identification, evidence collection, stance identification, and response generation~\cite{wang2024factcheck}. As showed in Fig~\ref{fig:system-flow}, this scaffolding preserves human agency by enabling contributors to iteratively refine outputs at each stage or perform high-level audits of the final results (DG3). To facilitate verification, CANote prioritizes transparency through provenance. Unlike systems relying on multi-hop~\cite{atanasova2024multi} or chain-of-thought~\cite{pan2023fact} reasoning, our pipeline explicitly links each subclaim to its corresponding evidence and note text (DG2). This granular traceability fulfills established transparency requirements~\cite{warren2025show}, grounding human-AI collaboration in verifiable evidence.

\textbf{Subclaim identification module.} Fact-checkers explicitly request AI assistance to disambiguate complex claims into verifiable units~\cite{liu2024human}. Accordingly, CANote uses LLMs to decompose original posts into succinct, standalone, and atomic subclaims, effectively scaffolding the initiation of the workflow (DG1). To optimize the user experience and manage computational costs, LLMs partition subclaims by punctuation~\cite{warren2025show,hu2025decomposition}, capped at four subclaims per post to prevent cognitive overload. Users retain full agency to verify, edit, add or delete subclaims before proceeding (DG3). We avoided importance ranking or feature scores to mitigate potential bias~\cite{ribeiro2016should}.

\textbf{Evidence collection module.} Fact-checkers exhibit algorithm aversion toward simple veracity predictions or opaque confidence scores, preferring transparent rationales and explicit source provenance~\cite{liu2024human}. To retrieve comprehensive results, CANote leverages the gemini-3.1-pro-search and gpt-5.1-search models, favoring LLM-driven searches over standard APIs (e.g., SerpAPI) to preserve context for complex claims. Through structured prompts, CANote transparently links retrieved evidence to corresponding subclaims, highlighting concise, phrase-level rationales within the original text (DG2). Evidence is deduplicated by matching exact titles and URLs across both host and path. To maintain focus, CANote retrieves a minimum of five URLs per request but caps the final deduplicated presentation at eight sources, ranked by claim relevance. Users can manually adjust the evidence stance via a dropdown menu, and only user-selected sources advance to the final synthesis stage (DG3). To help fact-checkers counteract confirmation bias and maintain neutrality~\cite{liu2024human}, the module intentionally retrieves and categorizes evidence representing both supporting and opposing stances (DG3). Each piece of evidence is paired with succinct, natural language explanations and direct quotes showing its correlation to the subclaim. We prioritize these short rationales over lengthy explanations~\cite{jacovi2020towards,liao2021human,warren2025show}, attention highlights~\cite{popat2018declare}, or saliency maps~\cite{petsiuk2021black,adebayo2018sanity,bibal2022attention,wiegreffe2019attention} which are often less reliable or legible. 

\textbf{Stance identification module.} CANote evaluates the stance of each evidence piece relative to a subclaim using the ``ibm-research/ claim\_stance'' model from Hugging Face. The identified results are shown alongside the evidence, and color-coded as red (Oppose) or green (Support). To overcome algorithm version and provide the transparent rationale that fact-checkers usually expect~\cite{liu2024human}, CANote provides direct provenance through interactive visual linkages (DG2). We embed subclaims and evidence using SentenceBERT to compute similarity scores, dynamically ranking and displaying relevant evidence when a subclaim is selected, and vice versa. All entities (subclaims, evidence, and debunking note text) are assigned unique IDs, allowing the LLMs to accurately link them. We use a URL parser to prevent link hallucinations. We omit numerical confidence metrics, as they are often confusing~\cite{warren2025show} and unreliable due to LLMs' tendency to overestimate confidence~\cite{tanneru2024quantifying,steyvers2025large}. Instead, presenting contradicting opinions naturally signals uncertainty and ensures impartiality~\cite{kotonya2020explainable} (DG3). We abstain from forcing a single definitive label, aligning with the perspective that AI is better suited for low-stakes assistance than high-stakes verdicts~\cite{ashoori2019ai,juneja2022human}.

\textbf{Response generation module.} Prior work found that fact-checkers benefit from AI support in generating structured drafts for their reports~\cite{liu2024human}. To achieve this without the over-interpretation risks (DG1), CANote concatenates the user-validated subclaims, reasoning and evidence into a cohesive, neutral debunking note~\cite{liu2024human}. Because explainable fact-checking must exceed simple token highlighting~\cite{popat2018declare} or basic text summaries~\cite{atanasova2020generating,kotonya2020explainable}, we show correlated evidence and subclaims as cite markers within each debunking note sentence (DG2). Users retain complete control to refine the text, manage URLs, or modify content at the subclaim level (DG3).

\subsection{Implementation Details}

CANote is implemented as a web-based technique for social media platforms. The technique maintains a structured, state-driven workflow comprising subclaim extraction, source curation, note modification, and final source selection.

\textbf{Data architecture.} At the core of CANote, each drafting session is formalized as a tuple: (sentences, evidence, links, selections). Sentences store claim-level text. Evidence encapsulates normalized source objects (URL, title, rationale, stance, and trust metadata). Links define many-to-many mappings between sentence indices and evidence IDs. Selections record the user's inclusion or exclusion decisions. This index-based, persistent linkage allows for granular, sentence-level editing (e.g., keep, rewrite, or delete) while ensuring that interaction edits remain synchronized across workflow steps.

\textbf{Backend processing pipeline.} For evidence retrieval, CANote specifically uses search-enabled LLMs to search for potential sources. The automated pipeline sequentially performs source aggregation, URL normalization, deduplication, and stance labeling (i.e., support or oppose). CANote then computes relevance alignments between the retrieved evidence text and subclaim-level text. These alignments initialize the link indices and determine the default ordering of evidence. Note generation uses gemini-3.1-pro, the most advanced model.

\section{User Study: Evaluating CANote}

To empirically evaluate the efficacy of our AI-assisted contribution technique, we designed a user study. The primary objective was to assess how the integration of provenance-based evidence retrieval and scaffolded co-drafting impacts authoring efficiency, note quality, and the contributors' perceived cognitive load and sense of agency. Beyond fact-checkers, we also sought to validate whether lay users could write high-quality notes with CANote. 

\subsection{Study Design}

\begin{figure}[!htbp]
    \includegraphics[width=0.47\textwidth]{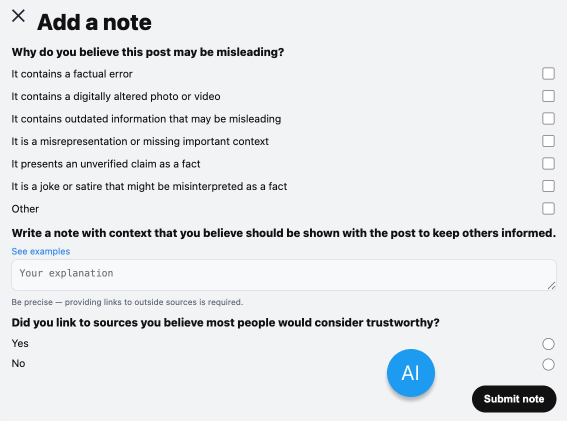}
    \caption{Experimental interface of the baseline technique.}
    \label{fig:interface}
\end{figure}

We employed a between-subjects design to mitigate learning effects, carryover biases, and cognitive fatigue. Because fact-checking involves analytical strategies, prior exposure to AI-assisted scaffolding in CANote could alter users' verification workflows and expectations. We chose two conditions to evaluate CANote's effectiveness compared to traditional debunking note drafting methods, as in Fig~\ref{fig:interface}:

$\bullet$ Baseline: A standard interface mimicking the current $\mathbb{X}$'s Community Notes workflow, where participants manually search for evidence using their own web browser and draft notes from scratch. To isolate the effects of CANote's scaffolding from general LLM capabilities, we provided an AI assistant (with gemini-3.1-pro backend, search enabled) on the interface.  

$\bullet$ CANote: As in Sec~\ref{sec:CANote}, it features the four-stage pipeline for subclaim identification, evidence collection, stance identification and note generation. 

We acquired recent English posts at the time of the study (March 1st, 2026 to March 15th, 2026) through the official $\mathbb{X}$ AI Note Writer API, to ensure participants had not previously encountered the corresponding posts, and to improve ecological validity. These posts were those for which users had requested notes on $\mathbb{X}$, mimicking the setting of the current $\mathbb{X}$'s note-writing system. We only retrieved English posts as we targeted UK and US participants, which are also where several of $\mathbb{X}$'s official pilot tests took place. These geographical and language choices facilitated comparison with $\mathbf{X}$'s official mechanisms. Both systems received the same feed of posts, and the feed of posts adopted an infinite-scroll interface design. We empirically observed that participants did not scroll to posts past the first 100 posts. To ensure experimental control, participants were notified that while the use of external web browsers was permitted, LLM-based chatbots and generative search engines were prohibited. We measured several objective and subjective dimensions:

$\bullet$ Objective metrics: We measured note writing time per note and note quality. Note quality was operationalized via the claim opinion score from the Community Note's AI Note evaluation criteria~\cite{x_community_notes_api}. To account for content safety, we also evaluated the notes for toxicity\footnote{Detoxify-original: https://huggingface.co/unitary/toxic-bert} and the probability of hateful speech\footnote{HateBERT: https://huggingface.co/Jensvollends/hatebert-finetuned\_v5} using two widely adopted fine-tuned models. For human evaluation, following prior work~\cite{chuai2025request}, we recruited 8 annotators with experience rating posts on $\mathbb{X}$ to assess note helpfulness. After being re-informed of $\mathbb{X}$'s official guidelines, they categorized all generated notes as \textit{helpful}, \textit{somewhat helpful}, or \textit{not helpful}. The annotation process yielded an inter-rater reliability of 0.65, indicating substantial agreement. To contextualize the writing quality, we also evaluated a commercial baseline currently deployed in $\mathbb{X}$'s Community Notes system, denoted as \textit{AI}.\footnote{Specifically, the collaborative note feature, and its open-sourced repository is available \href{https://github.com/twitter/communitynotes}{here}.} 

$\bullet$ Subjective metrics: We measured cognitive load using the NASA Task Load Index (NASA-TLX)~\cite{hart1988development} to evaluate whether CANote causes cognitive overload. We evaluated users' trust in CANote via the Trust in Automation Scale~\cite{jian2000foundations}, user experience via the User Experience Questionnaire Short Version (UEQ-S)~\cite{hinderks2017design}, and perceived agency during the co-drafting phase following prior practices~\cite{draxler2024ai,fu2023comparing}. Except for the NASA-TLX which used 21-point scales, all entries used 7-point scales.


\subsection{Procedure}

The study was conducted online, with participants recruited via Prolific and redirected to a custom, self-hosted experimental platform. Following informed consent, participants received a standardized introduction to both the Community Notes guidelines and their assigned system. Participants then authored notes for posts, acting as note contributors. While a minimum of two notes was required for the study, further contributions were permitted to allow for natural engagement. Finally, participants completed questionnaires to evaluate their experiences.

\subsection{Participants and Recruitment}

We conducted an a priori power analysis ($\alpha$=0.05, 1-$\beta$=0.80, Cohen's d=0.80), which determined a target sample size of 26 participants per condition. Consequently, across the two techniques, we recruited 52 fact-checkers and 52 lay users via prolific, evenly distributing 26 participants from each group to either the CANote or baseline condition. Inclusion criteria for both groups required an approval rate $\ge$ 95\%, and current residence in the US or UK. Fact-checkers were additionally required to hold verifications on the prolific platform as AI fact-checkers. The study was approved by our institution's IRB. Participants were compensated according to Prolific's guidelines: fact-checkers received £6.25 (M=27 minutes) and lay users received £2.25 (M=16 minutes). Both groups exhibited diverse demographic backgrounds, with 82.7\% (N=43) of fact-checkers and 76.9\% (N=40) of lay users holding at least a Bachelor's degree. The vast majority of participants were native English speakers (Fact-checker: N=49, Lay user: N=47). While fact-checkers inherently possessed professional expertise, a notable subset of lay users (N=20) reported prior experience in moderation or fact-checking tasks. Familiarity with $\mathbb{X}$'s Community Notes was low across both groups, while generative AI adoption was prevalent. Detailed demographic and background distributions are in Table~\ref{tab:demographics}.

\begin{table}[htbp]
\centering
\caption{Demographics and experience of participants.}
\label{tab:demographics}
\resizebox{\columnwidth}{!}{%
\begin{tabular}{llcc}
\toprule
\textbf{Category} & \textbf{Level} & \textbf{Fact-checker} & \textbf{Lay users} \\ \midrule
\textbf{Gender} & Male & 26 & 29 \\
 & Female & 25 & 22 \\
 & Non-binary & 1 & 1 \\ \midrule
\textbf{Age} & 18--35 & 15 & 28 \\
 & 36--55 & 33 & 20 \\
 & 56+ & 4 & 4 \\ \midrule
\textbf{Race / Ethnicity\textsuperscript{*}} & White or Caucasian & 38 & 33 \\
 & Asian & 5 & 10 \\
 & Black or African American & 6 & 6 \\
 & Hispanic or Latino & 5 & 1 \\
 & Middle Eastern or North African & 2 & 0 \\
 & Another race or ethnicity & 0 & 1 \\ \midrule
\textbf{Education} & Doctorate or Professional degree & 5 & 3 \\
 & Master's degree & 17 & 11 \\
 & Bachelor's degree & 21 & 26 \\
 & Some college or High school & 9 & 12 \\ \midrule
\textbf{Prior Experience\textsuperscript{*}} & Professional Fact-checking & 10 & 7 \\
 & Community or Content Moderation & 16 & 18 \\
 & Social Media Explanatory Notes & 3 & 9 \\ \bottomrule
\end{tabular}%
}%
\end{table}

\subsection{Data Analysis}

For quantitative analysis, objective task completion times and subjective questionnaire scores were analyzed using Welch's t-tests, or Mann-Whitney U tests for non-normally distributed data, to determine statistical significance. For comparisons involving more than two conditions, we conducted Welch's one-way Analysis of Variance (ANOVA). Audio recordings and open-ended survey responses from the exit interviews were transcribed and analyzed using thematic analysis. Two experimenters inductively coded the transcripts to identify recurring themes, and met intermittently to resolve any disagreements. We did not calculate inter-rater reliability as it may adversely affect the coding results~\cite{mcdonald2019reliability}. When reporting the results, we used ``NXX'' to denote lay users and ``EXX'' to denote fact-checkers (i.e., experts).

\subsection{Results}

\subsubsection{Note Writing Quality and Time}

To evaluate note-writing performance, we benchmarked \textit{CAnote} and \textit{baseline} against the fully automated \textit{AI} condition. We observed a significant main effect of technique on note quality ($F_{2, 28.13} = 59.92$, $p < .001$, $\eta^2_p = .810$). Post-hoc comparisons revealed that both \textit{CANote} ($M = 0.09$) and \textit{AI} ($M = 0.08$) significantly outperformed the \textit{baseline} ($M = -1.71$, adjusted $p < .001$), while no significant difference was found between \textit{CANote} and \textit{AI} ($p = .958$). This suggests that \textit{CANote} effectively helped users to generate high-quality notes.

We did not find significant differences in writing times between \textit{CANote} and the \textit{baseline} (Welch's $t = 5.96$, $p = .629$). Finally, we identified a significant effect of technique on perceived helpfulness ($F_{2, 27.10} = 42.80$, $p < .001$, $\eta^2_p = .760$). Both \textit{CANote} (M=0.68) and \textit{AI} (M=0.71) were rated as significantly more helpful than the \textit{baseline} (M=0.29, adjusted $p < .001$ and $p = .015$, respectively), with no significant difference between the two ($p = .851$). Collectively, these results suggest that while \textit{CANote} does not significantly reduce temporal costs compared to manual writing, it substantially enhances the utility and quality of the debunking notes.



In contrast, note-writing time was statistically comparable between conditions (\textit{CANote}: M=5.33 min; \textit{baseline}: M=5.96 min; $\Delta=-0.63$ min), with no significant difference (Welch's t=-0.413, p=.680) and negligible effect size (Cohen's d=.061).

\begin{figure}[!htbp]
    \centering 
    \includegraphics[width=0.5\textwidth]{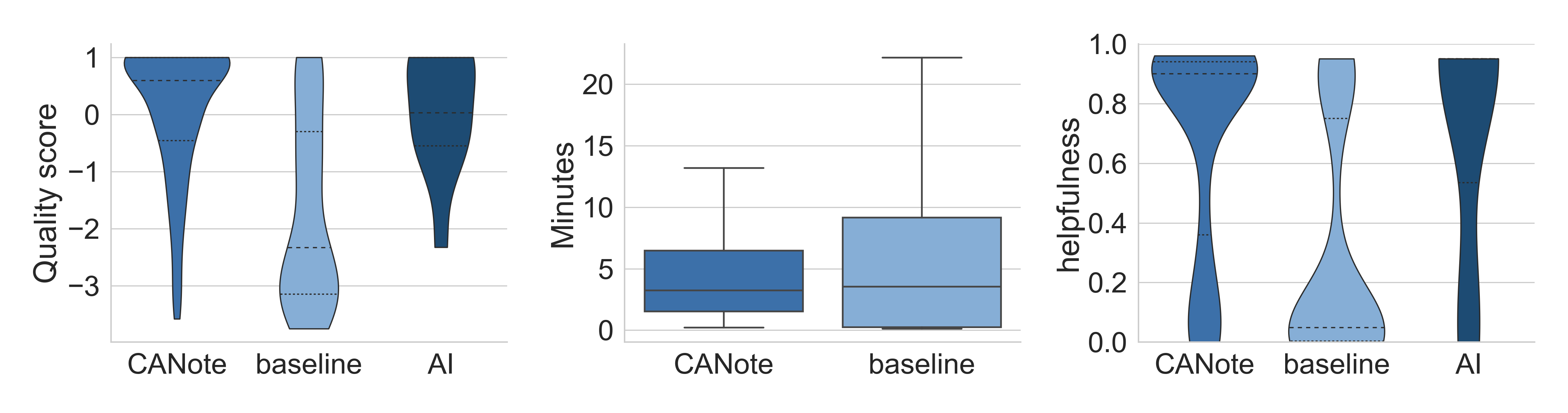}
    \caption{Note quality (left), note writing time (middle) and helpfulness score (right) for CANote, baseline and the AI system.}
    \label{fig:quality_time}
\end{figure}

\begin{figure}[!htbp]
    \centering 
    \includegraphics[width=0.5\textwidth]{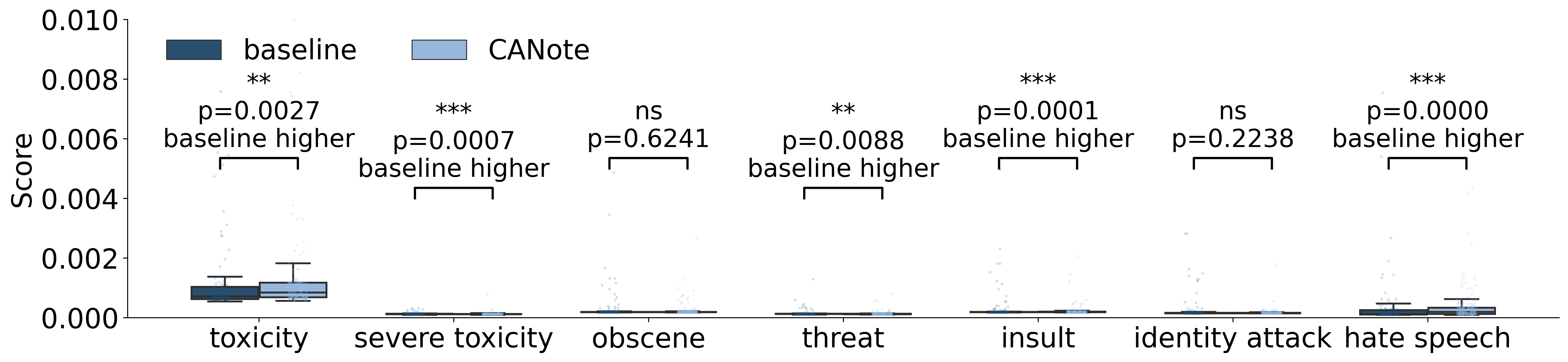}
    \caption{Toxicity and hateful-speech probabilities of notes across different techniques (lower is better).}
    \label{fig:toxicity}
\end{figure}

We finally evaluated notes' toxicity and hate-speech probability (Figure~\ref{fig:toxicity}). Overall, both baseline and CANote notes remained at low risk levels, indicating that unsafe content was uncommon in both conditions. Besides, CANote showed lower scores on most toxicity-related dimensions and hate-speech probability. These indicate that CANote support users in writing notes while maintaining low levels of sensitive or harmful language.

\subsubsection{Behavioral Patterns}\label{sec:behavioral}

We further analyzed how users interacted with CANote. Table~\ref{tab:data_summary} shows the frequencies of modifying or deleting sub-claims, modifying sources, and editing notes at both the sub-claim and character levels.

\begin{table}[ht]
\centering
\caption{Behavioral patterns for CANote, see Appendix~\ref{sec:definition} for detailed definitions.}
\label{tab:data_summary}
\resizebox{0.5\textwidth}{!}{%
\begin{tabular}{lrlr} 
\toprule
Metric & Mean (SE) & Metric & Mean (SE) \\
\midrule
\textbf{Evidence Interactions} & & \textbf{Actions} & \\
toggle\_evidence & 3.74 (0.74) & edited\_note & 76.40\% (5.01\%) \\
toggle\_supporting\_source & 1.18 (0.28) & edited\_subclaim & 29.20\% (5.36\%) \\
apply\_sources & 0.82 (0.14) & selected\_source & 51.40\% (5.89\%) \\
change\_evidence\_stance & 1.29 (0.34) & changed\_stance & 27.80\% (5.28\%) \\
evidence\_link\_clicked & 1.85 (0.62) & & \\
\midrule
\textbf{Editing Events} & & \textbf{Content Delta} & \\
claim\_edit\_events & 1.46 (0.42) & source number  & 3.54 (0.13) \\
note\_input\_events & 23.60 (3.28) & 1st $\to$ 2nd length & -19.60 (15.10) \\
& & 1st $\to$ 2nd source & 0.17 (0.16) \\
\midrule
\textbf{Stance Shifts} & & & \\
oppose $\to$ support & 46.00 (6.78) & support $\to$ oppose & 37.00 (6.08) \\
\bottomrule
\end{tabular}
}%
\end{table}

Most participants (76.40\%) manually edited the generated notes, averaging 23.60 input events per session. This suggests that users used AI drafts as a structural scaffold. Similarly, 51.40\% actively curated the evidence by selecting specific sources, indicating that they maintained a high level of agency during synthesis. Frequent interaction with transparent UI elements, including 3.74 evidence toggles and 1.85 links clicks per interaction, shows that users actively traced the relationships between subclaims and external sources. This transparency reallocates effort toward verification and mitigates automation bias.

Users also scrutinized AI outputs. 27.80\% of users modified the AI-annotated evidence stances. Stance shifts in both directions, specifically 46 from oppose to support and 37 from support to oppose, confirm that participants exercised independent critical judgment rather than following system suggestions. Furthermore, 29.20\% of users edited subclaims, showing the necessity of human oversight for claim decomposition. Finally, the second notes averaged 19.60 characters shorter than the first, suggesting that users synthesized information more concisely as they gained familiarity with the system scaffolding.

\subsubsection{Subjective Ratings}\label{sec:subjective}

\begin{figure}[!htbp]
    \centering 
    \includegraphics[width=0.5\textwidth]{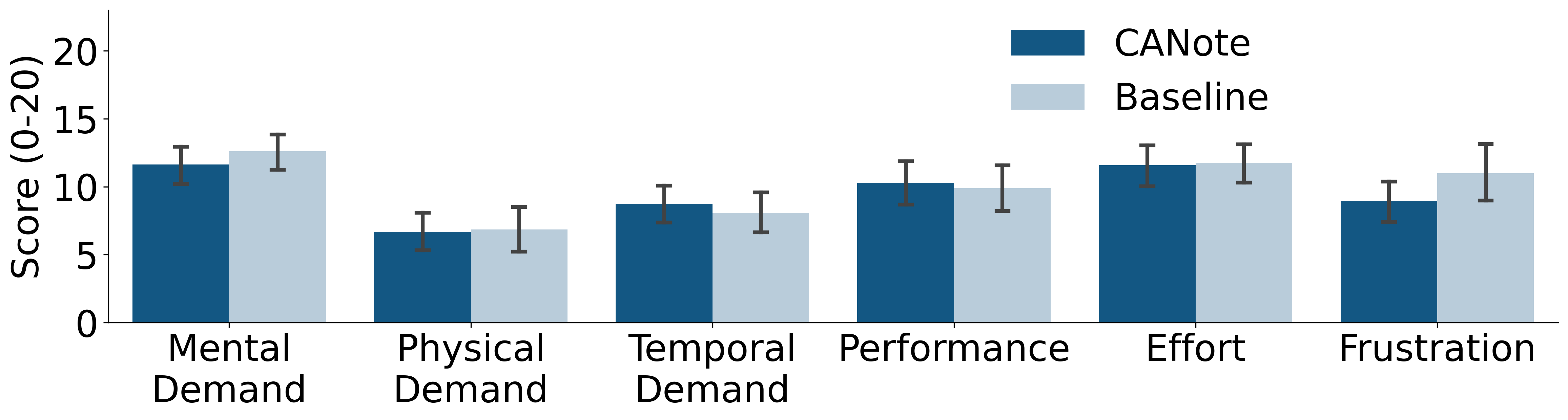}
    \caption{NASA-TLX ratings across techniques (0=best performance, 20=worst performance). }
    \label{fig:nasa_tlx}
\end{figure}

As showed in Fig~\ref{fig:nasa_tlx}, participants reported comparable subjective workloads across both techniques. Interestingly, we found no significant differences ($p > .05$) between the two techniques across any of the six NASA-TLX dimensions (\textit{Mental Demand}, \textit{Physical Demand}, \textit{Temporal Demand}, \textit{Performance}, \textit{Effort}, \textit{Frustration}). This indicates that structured scaffolding did not increase or decrease the perceived workload of the task. 

\begin{figure}[!htbp]
    \centering 
    \includegraphics[width=0.5\textwidth]{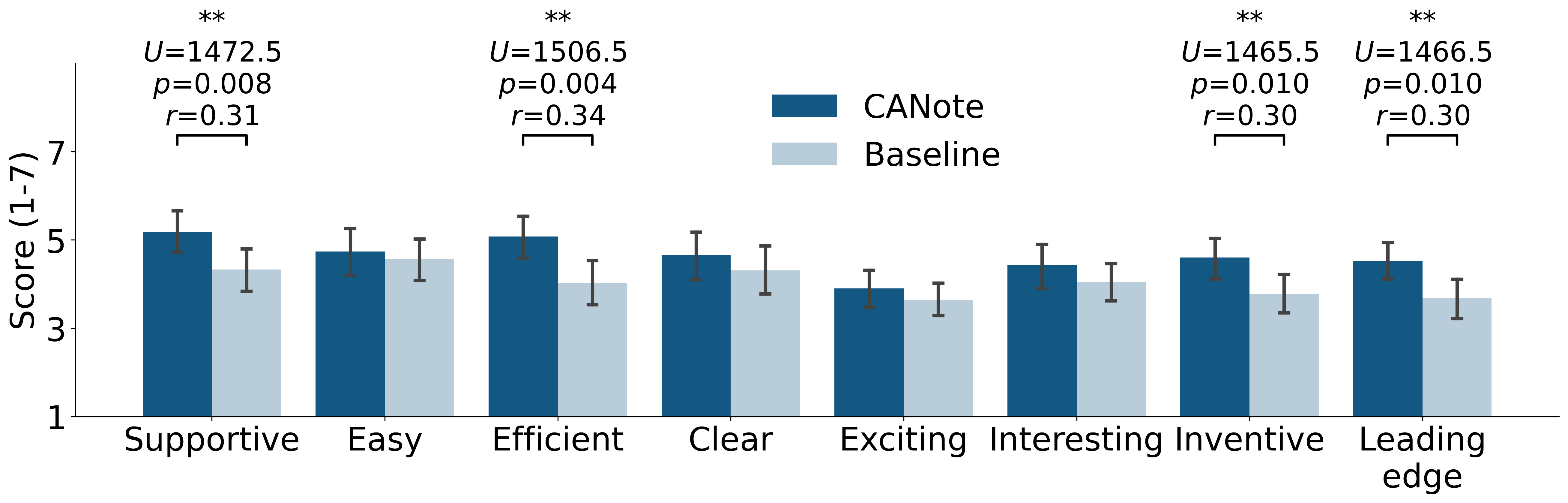}
    \caption{User experience ratings across techniques (1=most negative, 7=most positive). *, **, *** indicate $p < .05$, $< .01$ and $< .001$, respectively.}
    \label{fig:ueq}
\end{figure}

The UEQ ratings (Fig~\ref{fig:ueq}) reveal that CANote significantly improves several user experience dimensions. Specifically, participants rated CANote as significantly more \textit{Supportive} ($p = .008 < .01$), \textit{Efficient} ($p = .004 < .01$), \textit{Inventive} ($p = .010 < .05$) and \textit{Leading edge} ($p = .010 < .05$) than the baseline technique. This suggests that CANote's scaffolding effectively fosters a streamlined and creative authoring process.

Regarding trust, participants perceived CANote as significantly more \textit{Reliable} ($U = 1401.5$, $p = .037 < .05$, $r = .25$). However, we found no significant difference in the overall trust scores ($p = .469$) or across negative distrust items, such as being deceptive or suspicious. This likely reflects the double-edged nature of transparency. While valid evidence enhances trust, the visible exposure of failures facilitates critical scrutiny and prevents over-reliance.

\begin{figure}[!htbp]
    \centering 
    \includegraphics[width=0.5\textwidth]{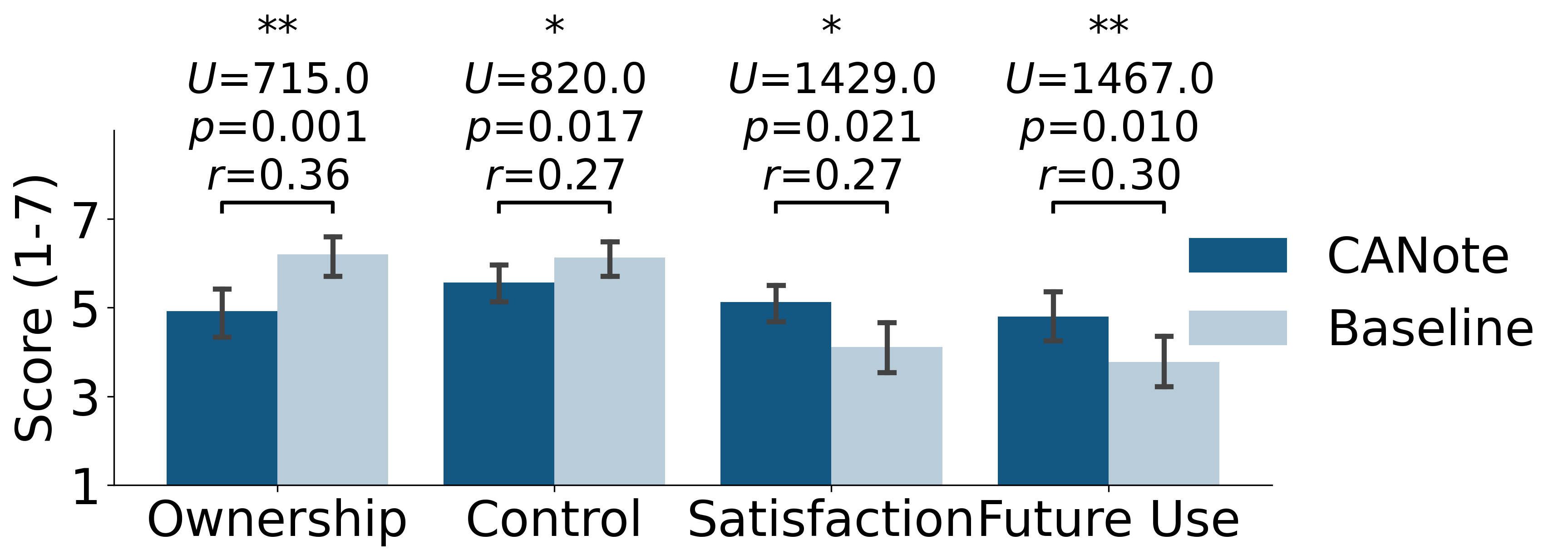}
    \caption{Different rating dimensions across different techniques (1=most negative, 7=most positive). *, **, *** indicate $p < .05$, $< .01$, $< .001$, respectively.}
    \label{fig:tech_control}
\end{figure}

Finally, Fig~\ref{fig:tech_control} highlights a trade-off between satisfaction and autonomy. While CANote significantly increased overall \textit{Satisfaction} ($p = .021 < .05$) and users' desire for \textit{Future use} ($p = .010 < .05$), it negatively impacted user autonomy. Participants using CANote reported a significantly lower sense of \textit{Ownership} over their written debunking notes ($p = .001 < .01$), and felt they had significantly less \textit{Control} over the writing process ($p = .017 < .05$). 

\subsubsection{Impact of User Expertise and Demographics}\label{sec:expertise}

To evaluate whether CANote's effectiveness differ across participant expertise, we compared the performance of lay users and experts. In the baseline condition, experts produced significantly higher-quality notes than lay users (mean $\Delta=1.201$, Welch's $t=2.859$, $p = .007 < .01$, Hedge's $g= 0.876$). Notably, CANote effectively bridged this performance gap. Lay users using CANote produced notes comparable in quality to those of experts using CANote (mean $\Delta=0.294$, Welch's $t=0.778$, $p=.443$, Hedge's $g=0.256$). Remarkably, lay users assisted by CANote significantly outperformed the unassisted expert group ($t = 4.374$, $p < .001$, Hedge's $g = 1.379$). 

Beyond performance, usability perceptions varied significantly by expertise. Experts rated CANote higher across multiple dimensions, finding it more \textit{Supportive} ($U = 1472.5$, $p = .008 < .01$, $r = .31$), \textit{Efficient} ($U = 1506.5$, $p = .004 < .01$, $r = .34$), \textit{Incentive} ($U = 1465.5$, $p = .010 < .05$, $r= .30$), and \textit{Leading edge} ($U = 1466.5$, $p = .010 < .05$, $r = .30$). Similarly, among trust dimensions, experts found CANote more \textit{Reliable} ($U = 1401.5$, $p = .037 < .05$, $r =.25$), though no significant differences are observed in negative trust evaluations (e.g., deceptive, harmful). These results suggest that expertise fosters a better-calibrated mental model of the system's operational boundaries, enabling experts to leverage its capabilities with greater confidence, even when their relative performance gains are smaller than those of lay users. 

Interestingly, we observed a paradoxical trade-off regarding user autonomy. While experts reported significantly higher overall \textit{Satisfaction} ($U = 1429.0$, $p = .021 < .05$, $r = .27$) and a stronger intent for \textit{Future use} ($U = 1467.0$, $p = .010 < .05$, $r = .30$), they experienced a significantly lower sense of \textit{Ownership} over the final debunking Note ($U = 715.0$, $p = .001 < .01$, $r = .36$), and perceived less \textit{Control} over the writing process ($U = 820.0$, $p = .017 < .05$, $r = .27$) compared to lay users. This implies that as users become more adept with the system, they may willingly defer to its automated structural interventions, trading subjective authorship for heightened efficiency and output quality.

Despite these divergences, there were no significant differences across six NASA-TLX workload dimensions (e.g., \textit{Mental Demand}, \textit{Frustration}, \textit{Effort}) between the two groups. This indicates that CANote remains highly usable, allowing lay users to utilize the tool effectively without suffering from cognitive overload.


Finally, our analysis stratified participants by age into younger ($<$45 years) and older ($\ge$45 years) groups to examine potential demographic variances. Older adults exhibited a disproportionate cognitive barrier, reporting higher \textit{Mental Demand} ($U = 1204.5$, $p = .029 < .05$, $r = .28$), \textit{Effort} ($U = 1228.5$, $p = .018 < .05$, $r = .31$), and \textit{Frustration} ($U = 1231.0$, $p = .017 < .05$, $r = .31$). In contrast, younger users found the note-writing task significantly more \textit{Interesting} ($U = 697.5$, $p = .046 < .05$, $r = .26$). The absence of significant age-based differences in other usability dimensions suggests that while CANote is functionally viable for all users, the cognitive friction of integrating CANote into the workflow is age-sensitive, thereby impacting engagement for older adults.

\subsubsection{Subjective Feedback}\label{sec:feedback}

We first analyzed the subjective feedback of the collaboration between humans and CANote, and then presented feedback on specific system features.

\textbf{\textit{Collaboration patterns.}} \textit{Workflow scaffolding.} Rather than relying on AI solely for content generation, participants used CANote to organize their analytical processes and manage their cognitive load. The AI functioned as a structural framework, parsing complex information into verifiable segments. Emphasizing this, E26 stated, \textit{``I used AI suggestions to break down the claims into smaller, checkable subclaims and guide the evidence search process.''} Similarly, prioritizing the breakdown of complex posts, N26 found it highly beneficial \textit{``to break down the claims into smaller, checkable subclaims.''}

\textit{Co-auditing outputs.} Many participants adopted a co-auditing strategy, leveraging the system to gather initial sources while rigorously verifying the AI's logic and outputs. Users manually traced the links provided by the AI to ensure factual alignment and source credibility. Highlighting this critical oversight, E40 noted, \textit{``I used AI to start the process and then manually went to every source [it] quoted to confirm if they were legit.''} E14 further detailed this verification process, stating, \textit{``I searched the site [the AI] used and read the articles to make sure the information matched. I also looked at the sites to see if they were reputable ...''} 

\textit{Independent corroboration.} Expert users frequently formulated independent mental models of the claim before consulting the AI, using the system to corroborate their interpretations. As E39 explained, \textit{``For the most part, then, the AI system was used to corroborate my own sense of what the post's claims were.''} While this human-led approach preserves analytical agency, it introduces the risk of selective exposure, wherein users might preferentially accept AI outputs that align with their preconceptions. However, participants actively mitigated this cognitive bias through deliberate vigilance. They emphasized that the system must serve as an objective analytical aid rather than an echo chamber, with P33 noting, \textit{``The AI needs to support the user, not replace them, and should be a tool to improve understanding, not seek to affirm user assumption.''} (P33)

\textit{Task delegation.} Conversely, a subset of users opted for full delegation, transferring the cognitive burden of synthesis and drafting entirely to the system. These participants found the generated outputs sufficient for submission without further intervention. Showing this reliance, E45 remarked, \textit{``It basically wrote a comprehensive note in clear language, which saved me the time ... [it] did not need editing as it covered all the points necessary.''}

\textbf{\textit{System feedback.}} \textit{Cognitive efficiency.} Regarding system functionality, users reported that CANote mitigated writer's block and enhanced synthesis efficiency by providing neutral, well-structured drafts. The AI effectively served as cognitive support for drafting notes. N9 confirmed using the system to \textit{``make the note more professional in tone.''} Echoing this sentiment, N7 shared, \textit{``The sentences written were clear, concise, and neutral in tone. It was nice to have those sentences generated and not have to think of them myself.''}

\textit{Preserving agency.} Despite the system's efficiency, participants maintained a strong sense of agency, using CANote as a guide rather than a definitive author. Users frequently tailored the generated text to align with their preferred phrasing or explicitly rejected suggestions that conflicted with their independent analysis. N10 summarized this dynamic, stating, \textit{``I mainly followed the wording and adjusted it slightly to match my own tone and ensure it fit the instructions.''} Some participants held a more conservative stance, with N24 saying, \textit{``I preferred my own fact-checking.''} 

\textit{Source reliability.} Users were found scrutinizing the perceived credibility of the AI-retrieved sources and the provided links. Participants exhibited a willingness to disagree with the system when sources appeared questionable or mismatched the assigned stance. N18 highlighted this discrepancy, observing that \textit{``some sources were from non-reputable sources, or did not support or oppose the claim as the AI labeled.''} E8 similarly emphasized conditional reliance, noting, \textit{``I used the AI suggestions after verifying them for accuracy.''}

\textit{Output limitations.} Participants also identified specific system limitations, frequently modifying or discarding irrelevant subclaims and erroneous inferences generated by CANote. As E26 pointed out, \textit{``Some AI suggestions included vague or less relevant subclaims, such as emotional or non-factual parts of the post, which I chose to ignore.''} Furthermore, E39 highlighted occasional logic failures in CANote's subclaim extraction, stating, \textit{``The AI suggested that one of the claims was `Marco Rubio', I disagreed, because the claim was actually much likelier that Marco Rubio made the assertion ... So I just disregarded that specific claim suggestion.''}

\section{Discussions}

\subsection{Feasibility of CANote}

Our findings in Sec~\ref{sec:expertise} show that provenance-based, scaffolded AI assistance can substantially improve the quality of crowdsourced fact-checking without increasing time or perceived workload. Participants using CANote produced significantly higher-quality notes than those in the baseline, while task completion time and NASA-TLX scores remained comparable. Rather than reducing cognitive load, CANote reallocates it from effortful information foraging and synthesis to verification and editing, supported by frequent user interventions such as editing the generated text (76.4\%), curating sources (51.4\%), and adjusting stance labels (27.8\%). 

A key contribution of CANote lies in bridging expertise gaps: with CANote, lay users produced notes comparable in quality to those written by fact-checkers, eliminating the  performance differences observed in the manual condition. This indicates that CANote's structured pipeline, particularly subclaim decomposition and evidence-claim linkage, functions as procedural scaffolding, externalizing expert reasoning into the interface. In doing so, the system lowers the barrier to participation without requiring prior expertise, aligning with the broader goal of scaling crowdsourced moderation~\cite{li2025scaling}.

Interestingly, while users perceived CANote as significantly more reliable, their overall trust in the automation did not show a significant increase. Behavioral and subjective results suggest that this stability is a positive outcome of system transparency. By explicitly exposing evidence mappings, CANote fostered a critical co-auditing strategy rather than blind acceptance. Users actively scrutinized sources, traced links, and discarded erroneous AI-generated sentences, demonstrating sustained analytical agency.

This capability has important real-world implications. For emerging Community Notes-style programs on platforms such as YouTube~\footnote{https://blog.youtube/news-and-events/new-ways-to-offer-viewers-more-context/} and Meta~\footnote{https://www.meta.com/technologies/community-notes/}, AI-assisted systems could help overcome cold-start challenges by rapidly increasing contributor capacity and coverage. Even within mature systems like $\mathbb{X}$'s Community Notes, such tools can serve as onboarding and training mechanisms, helping novice users learn the structure of effective note-writing before contributing independently. More broadly, expanding participation through AI scaffolding may lead to faster response times, greater coverage, and more diverse perspectives, potentially mitigating the risks of bias and manipulation associated with automated approaches~\cite{truong2025community}. 

\subsection{Patterns of Human-AI Collaboration}

Human-AI collaboration is prone to overreliance~\cite{he2023knowing,he2023stated}. To mitigate this risk, CANote presents balanced evidence and links note texts to both supporting and opposing viewpoints, which users considered comprehensive and beneficial, as in Sec~\ref{sec:feedback}. This facilitates user deliberation by requiring engagement with conflicting information. While AI assistants often risk inducing alert fatigue, overtrust, or cognitive overload~\cite{warren2025show,liu2024human}, CANote maintained a stable perceived workload through structural organization of information, as in Sec~\ref{sec:subjective}. By only surfacing results upon explicit user intent and providing contrasting evidence, the system encourages critical thinking and reduces false alerts.

Although CANote scaffolds the evidence search process, it preserves user proactivity by presenting contrasting perspectives rather than steering intent~\cite{zhang2025exploring}. Despite the risk of automation bias~\cite{romeo2026exploring, schemmer2022influence}, our behavioral observations in Sec~\ref{sec:behavioral} indicate that users actively modified the generated text rather than accepting it without verification. While providing explanations can occasionally increase automation bias~\cite{romeo2026exploring,schemmer2022influence}, visualizing the underlying logic and causality serves to mitigate algorithmic bias~\cite{truong2025community}. 

Our findings in Sec~\ref{sec:behavioral} further reveal both proactive and reactive collaboration patterns. Proactive users utilized AI-generated evidence to substantiate their independent reasoning, while reactive users relied more heavily on the initial drafts and intervened only when necessary. This flexibility indicates that CANote supports co-drafting across varying degrees of human agency, ensuring that the system remains effective regardless of a user's fact-check style.

\subsection{Computational Latency and Cost}

While multi-agent frameworks offer advanced capabilities, they often incur substantial computational and financial overhead. Empirically, CANote exhibits a mean end-to-end generation latency of 54.73 seconds (median: 23.81 seconds) and processes an average of 77,697 tokens (55,014 input and 22,683 output), resulting in an estimated cost of \$0.38 per generation using the gemini-3.1-pro model. These resource demands can be mitigated by substituting LLMs with task-specific models for low-complexity subtasks, implementing caching for redundant operations, or utilizing lightweight search APIs such as SerpAPI. Notably, participants did not perceive these delays, as processing was distributed across pipeline stages and notes were cached prior to the writing tasks, as reflected in the high usability ratings in Sec~\ref{sec:subjective}. Such asynchronous strategies suggest that CANote's computational requirements are acceptable for real-world deployment.

\subsection{Resilience Against Bias and Manipulation}

Our findings suggest that CANote shifts bias from individual judgment during information foraging toward interaction with structured evidence, creating both new safeguards and new risks.

First, human contributors frequently over-debunk, disproportionately focusing their efforts on criticizing content they subjectively perceive as false~\cite{jang2025too}. CANote partially mitigates this by reframing fact-checking as contextualization rather than correction. By default, CANote generates structured notes for the specific post,  providing neutral context for accurate claims alongside corrections for misleading ones. Empirically, as detailed in Sec~\ref{sec:subjective}, we observed that the system encourages users to ground their responses in retrieved evidence rather than subjective judgment.

At the same time, human fact-checking is susceptible to selective exposure and confirmation bias~\cite{freedman1965selective,oswald2004confirmation}, wherein users preferentially search for and integrate information that aligns with their pre-existing beliefs. Traditional manual workflows exacerbate this issue, as users may formulate skewed search queries~\cite{warren2025show,tong2025unite}. CANote mitigates this cognitive echo chamber by automatically retrieving diverse evidence and explicitly categorizing it into contrasting stances (e.g., supporting vs. opposing) prior to the drafting phase. This strategy forces users to confront contested information states. However, even in this setting, users may cherry-pick sources. Therefore, future longitudinal studies are necessary to quantify the cognitive biases caused by CANote.

Beyond cognitive biases, CANote must also contend with adversarial and system-level risks. Adversaries may artificially elevate disinformation or discredit legitimate claims through generative engine optimization~\cite{aggarwal2024geo} or search engine manipulation~\cite{dai2024neural}. CANote mitigates these risks by presenting diverse evidence, deduplicating sources, and anchoring the generated text to traceable evidence, requiring broader manipulation of the information pool rather than isolated attacks. Future work could explore semantic deduplication and weighting of sources to strengthen robustness.

Conversely, internal risks stem from the underlying LLMs, which are vulnerable to hallucinations~\cite{huang2025survey} and backdoor injections that may introduce political bias into the generated text~\cite{yang2024watch}. Such vulnerabilities threaten downstream accuracy~\cite{choi2024llm} and could pollute community ranking algorithms, causing them to upvote incorrect notes and produce inconsistent helpfulness scores~\cite{truong2025community}. CANote mitigates these risks through direct provenance, anchoring all generated text to specific traceable source URLs to facilitate human oversight. However, this mixed-initiative workflow complicates accountability. When a co-drafted note is inaccurate, attributing responsibility between humans and the AI becomes ambiguous. Addressing this risk needs clear platform-level regulations, such as mandating explicit labels for all AI-assisted contributions, similar to those on $\mathbb{X}$'s Collaborative Note~\cite{communitynotes_collaborative_notes}.

\section{Limitations and Future Work}

We acknowledge several limitations of this work. First, to increase ecological validity and comparability, our study's scope aligned with existing real-world fact-checking platforms. Specifically, we evaluated CANote on real-world posts from $\mathbb{X}$. As $\mathbb{X}$ pioneered the crowdsourced Community Notes program and provides active APIs for both collaborative and AI notes, this environment provided an ecologically valid commercial baseline for our evaluation. The current focus leaves cross-platform transferability (e.g., to Facebook or TikTok) for future exploration. Similarly, our participant recruitment was limited to users in the US and UK, interacting with English content. This mirrored $\mathbb{X}$ Community Note program's choice for pilot tests, though we note future multi-lingual adaptation is needed. Furthermore, we tasked participants with authoring a controlled number of notes. This avoids fatigue and aligns with common numbers of note writing~\cite{chuai2024community}, though future deployments are needed to examine learning effects, user fatigue, and selective exposure.

Second, our evaluation of notes did not involve $\mathbb{X}$'s official ranking algorithm. Implementing this algorithm requires a substantial volume of user ratings by live deployment on the platform or by recruiting a large pool of raters with verifiable backgrounds. We avoided Live testing to prevent contaminating the active fact-checking environment, and to mitigate biases associated with note-writing timestamps. Besides, recruiting raters with detailed profiles raises privacy concerns. 

Third, CANote integrates a variety of models, such as gemini-3.1-pro-search and gpt-5.1-search. While prior research suggests that LLMs tend to converge when provided with identical prompts~\cite{jiang2025representation}, its performance stability across models needs further validation. 

Finally, we omitted formal evidence evaluation and checkworthiness detection. The absence of an evidence assessment stage may introduce biases or incorporate unreliable secondary sources~\cite{akhtar2023multimodal,warren2025show}. Future research should integrate metrics for source credibility and media bias to improve the robustness of the verification process~\cite{schlichtkrull2024generating,baly2018predicting}. Besides, because not all social media content warrants formal intervention, future work should implement checkworthiness detection to identify specific posts needing debunking notes.

\section{Conclusion}

This paper proposes CANote, a technique for scaffolding the drafting of community notes. CANote decomposes the fact-checking process into interleaving steps to foster collaboration, providing provenance-based mappings between subclaims, evidence, and the debunking note. An online evaluation with 52 fact-checkers and 52 lay users show that CANote significantly improves note quality. With CANote, lay users can produce notes that match experts. While CANote maintains a consistent task completion time and cognitive workload compared to manual methods, it significantly increases user satisfaction and willingness for future use.

\bibliographystyle{ACM-Reference-Format}
\bibliography{main}

\appendix

\section{Ethics Considerations}

We acknowledge the ethics implications in this paper, and adhered to the Belmont report~\cite{beauchamp2008belmont} and the Menlo report~\cite{bailey2012menlo} during the studies. 

\textbf{Respect for persons.} All protocols in this study involving human subjects were approved by our university's Institutional Review Board (IRB). Participants provided informed consent and were introduced to both the system and Community Notes guidelines prior to participation.

As this study may contain misinformation or misleading content, and some of these contain sensitive content or abuse, before the study we highlighted in prolific recruitment that the study may contain misleading content or misinformation that made you discomfort. Besides, before the study we also informed users in the informed consent that some of the content may be misleading.

\textbf{Beneficence.} To mitigate risks associated with exposure to social media environments, the evaluation was conducted in a controlled, simulated environment rather than through live deployment or official platform APIs. This approach prevented the inadvertent spread of unverified claims and avoided contaminating active informational ecosystems. Furthermore, CANote addresses potential AI hallucinations and adversarial manipulation by enforcing transparent reasoning through the preservation of both supporting and opposing evidence. Participants were compensated according to Prolific guidelines, with fact-checkers and lay users receiving £6.25 and £2.25, respectively, based on average completion times.

\textbf{Justice.} Recruitment spanned both verified professional fact-checkers and lay users to assess fact-checking capabilities. Besides, we did not pose restrictions in recruitment regarding race, educational background and age, and our recruitment results are diverse in race, educational background and age.

\textbf{Respect for law and public interest.} To protect participant privacy and prevent unauthorized dissemination of personally identifiable information during the evaluation, all written notes were strictly anonymized before evaluation. We also avoided directly using $\mathbb{X}$ for testing, to avoid the potential pollution of the written notes to the public environment. Finally, given that the study involved potentially sensitive or misleading content, we explicitly disclosed these risks during recruitment on Prolific and again within the informed consent form. This ensured that participants were fully aware of potential discomfort before proceeding with the task.

\section{Generative AI Usage}

We transparently disclosed the usage of generative AI in this paper. We used Gemini-3.1-pro and GPT-5.2 for polishing the text, checking the grammars and improving writing fluency for the main text. We used Gemini-nano-banana for editing and improving the figures in the paper, although the final version of the figures are manually drawn by authors. Authors hold the ultimate responsibility for all the content of this paper. 

\section{Codebook for the Evaluation}

Table~\ref{tab:codebook} showed the codebook for the qualitative analysis in our evaluation study.

\begin{table*}[htbp]
\centering
\caption{Codebook for the qualitative analysis.}
\label{tab:codebook}
\small
\begin{tabular}{@{}llp{8cm}@{}}
\toprule
\textbf{Theme} & \textbf{Code} & \textbf{Description} \\ \midrule
\multirow{5}{*}{\shortstack[l]{Collaboration \\ Patterns}} 
& Workflow scaffolding & Using AI to decompose complex claims into atomic, verifiable subclaims to guide reasoning. \\
& Co-auditing & Iterative verification of AI-generated summaries and source credibility through manual link tracing. \\
& Corroboration & Using AI outputs to validate or cross-reference the user's pre-existing mental model of the claim. \\
& Strategic vigilance & Conscious effort to mitigate AI echo chambers by actively seeking contrasting viewpoints. \\
& Task Delegation & Transferring the cognitive burden of synthesis and drafting to the system for efficiency. \\ \midrule
\multirow{4}{*}{\shortstack[l]{System \\ Evaluations}} 
& Cognitive efficiency & Perception of AI as a tool to overcome writer's block and ensure a professional, neutral tone. \\
& Agency preservation & Active tailoring of generated text to align with personal, independent analysis. \\
& Source scrutiny & Critical evaluation and rejection of non-reputable sources or mismatched stance labels. \\
& Logic and relevance failures & Identification and rectification of irrelevant subclaims or logical errors in AI-driven extraction. \\ \bottomrule
\end{tabular}
\end{table*}

\section{Definitions of the metrics in behavioral pattern analysis}\label{sec:definition}

The metrics in Table~\ref{tab:data_summary} reflected the average frequency of specific UI interactions per session. 

$\bullet$ toggle\_evidence: The times (per note) participants expanded or collapsed evidence details to view supporting or opposing information.

$\bullet$ toggle\_supporting\_source: The frequency (per note) with which participants toggled the visibility of specific sources linked to a subclaim.

$\bullet$ apply\_sources: The times (per note) participants clicked the ``apply'' button to integrate selected sources into the note draft. 

$\bullet$ change\_evidence\_stance: The frequency (per note) of manual overrides where users changed the AI-assigned stance (support/oppose) of a specific piece of evidence.

$\bullet$ evidence\_link\_clicked: The total count of clicks on URLs to verify the source of the evidence per note.

$\bullet$ edited\_note: The proportion of notes where the user manually modified the text of the generated note.

$\bullet$ edited\_subclaim: The proportion of notes where the user edited the AI-generated subclaim text. 

$\bullet$ selected\_source: The proportion of notes where the user manually selected or deselected specific sources for citation.

$\bullet$ changed\_stance: The proportion of notes where the user modified at least one AI-annotated stance.

$\bullet$ claim\_edit\_events: The average character of subclaim's modification recorded per note.

$\bullet$ note\_input\_events: The average character of note's modification recorded per note.

$\bullet$ source number: The average total number of unique sources cited per submitted note. 

$\bullet$ 1st $\rightarrow$ 2nd length: The mean difference in character count between the first submitted note and the second submitted note for a participant.

$\bullet$ 1st $\rightarrow$ 2nd source: The mean change in the number of unique sources cited between the first submitted note and the second submitted note for a participant.

\end{document}